\documentclass[aps,pre,superscriptaddress,amsmath,preprint,showpacs]{revtex4}
\usepackage{amsmath}
\usepackage[dvips]{graphicx}
\begin{document}
\title{Quantum and classical dynamics of Langmuir wave packets}

\author{F. Haas}
\altaffiliation{Universidade do Vale do Rio dos Sinos - UNISINOS, Av. Unisinos 950, 93022-000, S\~ao Leopoldo, RS, Brazil}
\affiliation{Institut f\"ur Theoretische Physik IV, Ruhr-Universit\"at Bochum\\ D-44780 Bochum, Germany}

\author{P. K. Shukla}
\altaffiliation{Department of Physics, Ume\aa \, University, SE-901 87, Ume{\aa}, Sweden}
\altaffiliation{GOLP / Instituto de Plasmas e Fus\~ao Nuclear, Instituto Superior T\'ecnico, Universidade T\'ecnica de Lisboa, 1049-001 Lisboa, Portugal}
\altaffiliation{SUPA, Department of Physics, University of Strathclyde, Glasgow, G40NG, UK}
\altaffiliation{School of Physics, University of Kwazulu-Natal, Durban 4000, South Africa.}
\affiliation{Institut f\"ur Theoretische Physik IV, Ruhr-Universit\"at Bochum\\ D-44780 Bochum, Germany}

\begin{abstract}
\noindent
The quantum Zakharov system in three-spatial dimensions and an associated Lagrangian description, 
as well as its basic conservation laws are derived. In the adiabatic and semiclassical case, 
the quantum Zakharov system reduces to a quantum modified vector nonlinear Schr\"odinger (NLS) 
equation for the envelope electric field. The Lagrangian structure for the resulting vector 
NLS equation is used to investigate the time-dependence of the Gaussian shaped localized solutions, 
via the Rayleigh-Ritz variational method. The formal classical limit is considered in detail. 
The quantum corrections are shown to prevent the collapse of localized Langmuir envelope fields, 
in both two and three-spatial dimensions. Moreover, the quantum terms can produce an oscillatory 
behavior of the width of the approximate Gaussian solutions. The variational method is shown 
to preserve the essential conservation laws of the quantum modified vector NLS equation. 
\end{abstract}
\maketitle

\section{Introduction}
The Zakharov system \cite{Zakharov}, describing the coupling between Langmuir and ion-acoustic waves, 
is one of the basic plasma models, see Ref. \cite{Goldman, Thornhill} for reviews. Recently \cite{Garcia}, 
a quantum modified Zakharov system was derived, by means of the quantum plasma hydrodynamic 
model \cite{Haas}--\cite{HaasQMHD}. In this context, enhancement of the quantum effects was 
then shown {\it e. g.} to suppress the four-wave decay instability. Subsequently \cite{Marklund}, 
a kinetic treatment of the quantum Zakharov system has shown that the modulational instability 
growth rate can be increased in comparison to the classical case, for partially coherent 
Langmuir wave electric fields. Also \cite{Haasvar}, a variational formalism was obtained and 
used to study the radiation of localized structures described by the quantum Zakharov system. 
Bell shaped electric field envelopes of electron plasma oscillations in dense quantum plasmas 
obeying Fermi statistics were analyzed in Ref. \cite{Shukla}.  More mathematically-oriented works 
on the quantum Zakharov equations concern its Lie symmetry group \cite{Tang} and the derivation of 
exact solutions \cite{Abdou}--\cite{Yang}. Finally, there is evidence of hyperchaos in the reduced 
temporal dynamics arising from the quantum Zakharov equations \cite{Misra}. 

All these paper refer to quantum Zakharov equations in one-spatial-dimension only. In the present work,
we extend the quantum Zakharov system to fully three-dimensional space, allowing also for the magnetic field 
perturbation. In the classical case, both heuristic arguments and numerical simulations indicate that 
the ponderomotive force can produce finite-time collapse of Langmuir wave packets in two- or 
three-dimensions \cite{Goldman}, \cite{Zakharov2, Zakharov3}. This is in contrast to the one-dimensional case, 
whose solutions are smooth for all time. A dynamic rescaling method was used for the time-evolution of 
electrostatic self-similar and asymptotically self-similar solutions in two- and three-dimensions, 
respectively \cite{Landman}. Allowing for transverse fields shows that singular solutions of the 
resulting vector Zakharov equations are weakly anisotropic, for a large class of initial 
conditions \cite{Papanicolaou}. The electrostatic nonlinear collapse of Langmuir wave packets 
in the ionospheric and laboratory plasmas has been observed \cite{Dubois, Robinson}. Also, 
the collapse of Langmuir wave packets in beam plasma experiments verifies the basic concepts 
of strong Langmuir turbulence, as introduced by Zakharov \cite{Cheung}. The analysis of 
the coupled longitudinal and transverse modes in the classical strong Langmuir turbulence 
has been less studied \cite{Alinejad}--\cite{Li}, as well as the intrinsically magnetized case \cite{Pelletier}, 
which can lead to upper-hybrid wave collapse \cite{Stenflo}. 
Finally, Zakharov-like equations have been proposed for the electromagnetic wave collapse in a 
radiation background \cite{Marklund2}.

It is expected that the ponderomotive force causing the collapse of localized solutions in two- or three-space 
dimensions could be weakened by the inclusion of quantum effects, making the dynamics less violent. 
This conjecture is checked after establishing the quantum Zakharov system in higher-dimensional 
space and using its variational structure in association with a (Rayleigh-Ritz) trial function method.  

The manuscript is organized in the following fashion. In Section 2, the quantum Zakharov system in 
three-spatial-dimensions is derived by means of the usual two-time scale method applied to the 
fully 3D quantum hydrodynamic model. In Section 3, the 3D quantum Zakharov system is shown to be 
described by a Lagrangian formalism. The basic conservation laws are then also derived. 
When the density fluctuations are so slow in time so that an adiabatic approximation is possible, 
and treating the quantum term of the low-frequency equation as a perturbation, a quantum modified 
vector nonlinear Schr\"odinger equation for the envelope electric field is obtained. 
In Section 4, the variational structure is used to analyze the temporal dynamics of localized (Gaussian) 
solutions of this quantum NLS equation, through the Rayleigh-Ritz method, in two-spatial-dimensions. 
Section 5 follows the same strategy, extended to fully 3D space. Special attention is paid to the 
comparison between the classical and quantum cases, with considerable qualitative and quantitative 
differences. Section 6 contains conclusions.  

\section{Quantum Zakharov equations in $3+1$ dimensions}
The starting point for the derivation of the electromagnetic quantum Zakharov equations is the 
quantum hydrodynamic model for an electron-ion plasma, Equations (20)-(28) of Ref. \cite{HaasQMHD}. 
For the electron fluid pressure $p_e$, consider the equation of state for spin $1/2$ particles at zero temperature,
\begin{equation}
\label{e1}
p_e = \frac{3}{5}\,\frac{m_{e}v_{Fe}^2 \,n_{e}^{5/3}}{n_{0}^{2/3}} \,,
\end{equation}
where $m_e$ is the electron mass, $v_{Fe}$ is the Fermi electron thermal speed, $n_e$ is the electron number 
density and $n_0$ is the equilibrium particle number density both for electron and ions. The pressure and 
quantum effects (due to their larger mass) are neglected for the ions. Also due to the larger ion mass, 
it is possible to introduce a two-time scale decomposition, $n_e = n_0 + \delta n_s + \delta n_f$, 
$n_i = n_0 + \delta n_s$, ${\bf u}_e = \delta{\bf u}_s + \delta{\bf u}_f$, 
${\bf u}_i = \delta{\bf u}_s$, ${\bf E} = \delta{\bf E}_s + \delta{\bf E}_f$, ${\bf B} = \delta{\bf B}_f$, 
where the subscripts $s$ and $f$ refer to slowly and rapidly changing quantities, respectively. 
Also, ${\bf u}_e$ is the electron fluid velocity, $n_i$ the ion number density, ${\bf u}_i$ the ion fluid velocity, 
${\bf E}$ the electric field, and ${\bf B}$ the magnetic field. Notice that it is assumed that there 
is no slow contribution to the magnetic field, a restriction which allows to get ${\bf B} 
= (m_{e}/e)\,\nabla\times\delta{\bf u}_f$ (see Equation (2.21) of Ref. \cite{Thornhill}), 
where $-e$ is the electron charge. Including a slow contribution to the magnetic field could 
be an important improvement, but this is outside the scope of the present work. 

Following the usual approximations \cite{Thornhill, Garcia}, the quantum corrected 3D Zakharov equations read 
\begin{eqnarray}
\label{e2}
2i\omega_{pe}\frac{\partial{\bf\tilde{E}}}{\partial t} &-& c^2\, \nabla\times(\nabla\times{\bf\tilde{E}}) 
+ v_{Fe}^2 \nabla(\nabla\cdot{\bf\tilde{E}}) = \nonumber \\ &=& \frac{\delta n_s}{n_0} \,\omega_{pe}^2 \,{\bf\tilde{E}} 
+ \frac{\hbar^2}{4m_{e}^2}\nabla\left[\nabla^2 (\nabla\cdot{\bf\tilde{E}})\right]  \,,  \\ 
\label{e3}
\frac{\partial^2 \delta n_s}{\partial t^2} &-& c_{s}^2 \,\nabla^2 \delta n_s 
- \frac{\varepsilon_0}{4m_i}\nabla^2 (|{\bf\tilde{E}}|^2) + \frac{\hbar^2}{4m_e m_i} \,\nabla^4 \delta n_s = 0 \,.
\end{eqnarray}
Here ${\bf\tilde{E}}$ is the slowly varying envelope electric field defined via
\begin{equation}
{\bf E}_f = \frac{1}{2}\,({\bf\tilde{E}} \, e^{-i\omega_{pe}t} + {\bf\tilde{E}}^{*} \, e^{i\omega_{pe}t}) \,,
\end{equation}
where $\omega_{pe}$ is the electron plasma frequency. Also, in Eqs. (\ref{e2}--\ref{e3}) $c$ is the speed of 
light in vacuum, $\hbar$ the scaled Planck constant, $\varepsilon_0$ the vacuum permittivity and $m_i$ the ion mass. 
In addition, $c_{s}^2 = \kappa_B T_{Fe}/m_i \,,$ where $\kappa_B T_{Fe} = m_e v_{Fe}^2$. 
Therefore, $c_s$ is a Fermi ion-acoustic speed, with the Fermi temperature replacing 
the thermal temperature for the electrons. 

In comparison to the classical Zakharov system (see Eqs. (2.48a)--(2.48b) of Ref. \cite{Thornhill}), 
there is the inclusion of the extra dispersive terms proportional to $\hbar^2$ in Eqs. (\ref{e2})--(\ref{e3}). 
Other quantum difference is the presence of the Fermi speed instead of the thermal speed in the last term 
at the left hand side of Eq. (\ref{e2}). From the qualitative point of view, the terms proportional 
to $\hbar^2$ are responsible for extra dispersion which can avoid collapsing of Langmuir envelopes, 
at least in principle. This possibility is investigated in Sections 4 and 5. Finally, notice 
the non trivial form of the fourth order derivative term in Eq. (\ref{e2}). It is not simply 
proportional to $\nabla^{4} {\bf\tilde{E}}$ as could be wrongly guessed from the quantum Zakharov 
equations in $1+1$ dimensions, where there is a $\sim \partial^{4}{\bf\tilde{E}}/\partial x^4$ 
contribution \cite{Garcia}.

It is useful to consider the rescaling  
\begin{eqnarray}
\label{e4}
\bar{\bf r} &=& \frac{2\sqrt{\mu}\,\omega_{pe}\,{\bf r}}{v_{Fe}} \,, \quad \bar{t} = 2\,\mu\,\omega_{pe}t \,,  \\
n &=& \frac{\delta n_s}{4\mu n_0} \,, \quad {\bf\cal E} = \frac{e\,\tilde{\bf E}}{4\,\sqrt{\mu}\,m_{e}\omega_{pe}v_{Fe}}  \,,  \nonumber
\end{eqnarray}
where $\mu = m_{e}/m_{i}$. Then, dropping the bars in ${\bf r}, t$, we obtain
\begin{eqnarray}
\label{e5}
i\frac{\partial{\bf\cal E}}{\partial t} &-& \frac{c^2}{v_{Fe}^2} \nabla\times(\nabla\times{\bf\cal E}) 
+ \nabla(\nabla\cdot{\bf\cal E}) = \nonumber \\ &=& n \, {\bf\cal E} 
+ \Gamma\,\nabla\left[\nabla^2 (\nabla\cdot{\bf\cal E})\right]  \,,  \\
\label{e6}
\frac{\partial^2 n}{\partial t^2} &-&  \nabla^2 n - \nabla^2 (|{\bf\cal E}|^2) + \Gamma\, \nabla^4 n = 0 \,,  
\end{eqnarray}
where
\begin{equation}
\label{e7}
\Gamma = \frac{m_e}{m_i}\left(\frac{\hbar\,\omega_{pe}}{\kappa_{B}T_{Fe}}\right)^2 
\end{equation}
is a non-dimensional parameter associated with the quantum effects. Usually, it is an extremely small quantity, 
but it is nevertheless interesting to retain the $\sim \Gamma$ terms, specially for the collapse scenarios. 
The reason is not only due to a general theoretical motivation, but also because from some simple estimates 
one concludes that these terms become of the same order as some of other terms in Eqs. (\ref{e2})--(\ref{e3}) 
provided that the characteristic length $l$ for the spatial derivatives becomes as small as the mean 
inter-particle distance, $l \sim n_{0}^{-1/3}$. Of course, the Zakharov equations are not able to describe 
the late stages of the collapse, since they do not include dissipation, which is unavoidable for short scales. 
But even Landau damping would be irrelevant for a zero-temperature Fermi plasma, where the main influence 
comes from the Pauli pressure. In the left-hand side of Eq. (\ref{e5}), the $\nabla(\nabla\cdot{\bf\cal E})$ 
term is retained because the $\sim c^{2}/v_{Fe}^2$ transverse term disappears in the electrostatic approximation. 

In the adiabatic limit, neglecting $\partial^{2} n/\partial t^2$ in Eq. (\ref{e6}) and under appropriated 
boundary conditions, it follows that 
\begin{equation}
\label{n}
n = - |{\bf\cal E}|^2 + \Gamma\, \nabla^{2} n \,,
\end{equation}
When $\Gamma \neq 0$, it is not easy to directly express $n$ as a function of $|{\bf\cal E}|$ as in the classical case. 
Therefore, the adiabatic limit is not enough to derive a vector nonlinear Schr\"odinger equation, due to the 
coupling in Eq. (\ref{n}).  

\section{La\-gran\-gian struc\-tu\-re and con\-ser\-va\-tion $\!\!\!$ laws}

The quantum Zakharov equations (\ref{e5})--(\ref{e6}) can be described by the Lagrangian density
\begin{eqnarray}
{\cal L} &=& \frac{i}{2}\,\Bigl(\,{\bf\cal E}^{*}\cdot\frac{\partial{\bf\cal E}}{\partial t} 
- {\bf\cal E}\cdot\frac{\partial{\bf\cal E}^{*}}{\partial t}\,\Bigr) 
- \frac{c^2}{v_{Fe}^2} |\nabla\times {\bf\cal E}|^2 - |\nabla\cdot{\bf\cal E}|^2 
- \Gamma \,|\nabla(\nabla\cdot{\bf\cal E})|^2 \nonumber \\
\label{e8}
&+& n\,\Bigl(\,\frac{\partial\alpha}{\partial t} - |{\bf\cal E}|^2\,\Bigr) 
- \frac{1}{2}\,\Bigl(n^2 + \Gamma |\nabla n|^2 + |\nabla\alpha|^2\Bigr) \,,
\end{eqnarray}
where $n$, the auxiliary function $\alpha$ and the components of ${\bf\cal E}, {\bf\cal E}^{*}$ are 
regarded as independent fields. Remark: for the particular form (\ref{e8}) and for a generic field $\psi$, 
one computes the functional derivative as  
\begin{equation}
\label{e9}
\frac{\delta{\cal L}}{\delta\psi} = \frac{\partial{\cal L}}{\partial\psi} 
- \frac{\partial}{\partial r_i}\,\frac{\partial{\cal L}}{\partial\psi/\partial r_i} 
- \frac{\partial}{\partial t}\,\frac{\partial{\cal L}}{\partial\psi/\partial t} 
+ \frac{\partial^2}{\partial r_{i}\,\partial r_j}\,
\frac{\partial{\cal L}}{\partial^{2}\psi/\partial r_i \partial r_j} \,,
\end{equation}
using the summation convention and where $r_i$ are cartesian components.

Taking the functional derivatives with respect to $n$ and $\alpha$, we have 
\begin{equation}
\label{e10}
\frac{\partial\alpha}{\partial t} = n + |{\bf\cal E}|^2 - \Gamma\nabla^2 n ,
\end{equation}
and
\begin{equation}
\label{e11}
\frac{\partial n}{\partial t} = \nabla^2 \alpha \,,
\end{equation}
respectively. Eliminating $\alpha$ from Eqs. (\ref{e10}) and (\ref{e11}) we obtain the low frequency equation. 
In addition, the functional derivatives with respect to ${\bf\cal E}^{*}$ and ${\bf\cal E}$ produce the 
high-frequency equation and its complex conjugate. The present formalism is inspired by the Lagrangian 
formulation of the classical Zakharov equations \cite{Gibbons}. 

The quantum Zakharov equations admit as exact conserved quantities the ``number of plasmons" of the Langmuir field, 
\begin{equation}
\label{e12}
N = \int |{\bf\cal E}|^2 \,d{\bf r} \,,
\end{equation}
the linear momentum  (with components $P_i, \, i = x, y, z$), 
\begin{equation}
\label{e13}
P_i = \int \Bigl[\frac{i}{2} \left({\cal E}_j \,\frac{\partial{\cal E}^{*}_j}{\partial r_i} 
- {\cal E}^{*}_j \,\frac{\partial{\cal E}_j}{\partial r_i}\right) 
- n \,\frac{\partial\alpha}{\partial r_i} \Bigr] \,d{\bf r} 
\end{equation}
and the Hamiltonian,
\begin{eqnarray}
{\cal H} &=& \int \Bigl[n|{\bf\cal E}|^2 + \frac{c^2}{v_{Fe}^2}\, |\nabla\times{\bf\cal E}|^2 
+ |\nabla\cdot{\bf\cal E}|^2 + \Gamma\,|\nabla(\nabla\cdot{\bf\cal E})|^2 \nonumber \\ \label{e14} 
&+&  \frac{1}{2}\,\Bigl(n^2 + \Gamma |\nabla n|^2 + |\nabla\alpha|^2\Bigr) \Bigr] \,d{\bf r} \,.
\end{eqnarray}
Furthermore, there is also a preserved angular momenta functional, but it is not relevant in the present work. 
These four conserved quantities can be associated, through Noether's theorem, to the invariance of the action 
under gauge transformation, time translation, space translation and rotations, respectively. 
The conservation laws can be used {\it e. g.} to test the accuracy of numerical procedures. 
Also, observe that equations (\ref{e6}) and (\ref{n}) for the adiabatic limit are described 
by the same Lagrangian density (\ref{e8}). In this approximation, it suffices to set $\alpha \equiv 0$.

In addition to the adiabatic limit, Eq. (\ref{n}) can be further approximated to
\begin{equation}
\label{e15}
n = - |{\bf\cal{E}}|^2 - \Gamma \nabla^{2}(|{\bf\cal{E}}|^{2}) \,,
\end{equation}
assuming that the quantum term is a perturbation. In this way and using Eq. (\ref{e5}), 
a quantum modified vector nonlinear Schr\"odinger equation is derived 
\begin{eqnarray}
i\frac{\partial{\bf\cal E}}{\partial t} &+& \nabla(\nabla\cdot{\bf\cal E}) 
- \frac{c^2}{v_{Fe}^2} \nabla\times(\nabla\times{\bf\cal E}) + |{\bf\cal{E}}|^2 {\bf\cal{E}} = \nonumber \\ 
\label{e16} &=& 
\Gamma\nabla\left[\nabla^2 (\nabla\cdot{\bf\cal E})\right] -\Gamma \, {\bf\cal E} \nabla^{2}(|{\bf\cal{E}}|^{2}) \,. 
\end{eqnarray}

The appropriate Lagrangian density ${\cal L}_{ad,sc}$ for the semiclassical equation (\ref{e16}) is given by 
\begin{eqnarray}
{\cal L}_{ad,sc} &=& \frac{i}{2}\,\Bigl(\,{\bf\cal{E}}^{*}\cdot\frac{\partial{\bf\cal{E}}}{\partial t} 
- {\bf\cal{E}}\cdot\frac{\partial{\bf\cal{E}}^{*}}{\partial t}\,\Bigr) 
- \frac{c^2}{v_{Fe}^2}|\nabla\times{\bf\cal{E}}|^2 - |\nabla\cdot{\bf\cal{E}}|^2 \nonumber \\ 
\label{e17}
&-& \Gamma \,|\nabla(\nabla\cdot{\bf\cal{E}})|^2 
+ \frac{1}{2}\,|{\bf\cal{E}}|^4 - \frac{\Gamma}{2}\,\Bigl|\nabla[\,|{\bf\cal{E}}|^2] \Bigr|^2 \,,
\end{eqnarray}
where the independent fields are taken as ${\bf\cal{E}}$ and ${\bf\cal{E}}^{*}$ components. 

The expression $N$ for the number of plasmons in Eq. (\ref{e12}) remains valid as a constant of motion 
in the joint adiabatic and semiclassical limit, as well as the momentum ${\bf P}$ in 
Eq. (\ref{e13}) with $\alpha \equiv 0$. Finally, the Hamiltonian 
\begin{eqnarray}
{\cal H}_{ad,sc} = \int \Bigl[\,\frac{c^2}{v_{Fe}^2}\,|\nabla&\times&{\bf\cal{E}}|^2 
+ |\nabla\cdot{\bf\cal{E}}|^2 + \Gamma \,|\nabla(\nabla\cdot{\bf\cal{E}})|^2 \nonumber \\ \label{e18} 
&-&  \frac{1}{2}\,|{\bf\cal{E}}|^4 + \frac{\Gamma}{2}\,\Bigl|\nabla[\,|{\bf\cal{E}}|^2\,] \Bigr|^2 \,\,\Bigr] \,d{\bf r} 
\end{eqnarray}
is also a conserved quantity.   

In the following, the influence of the quantum terms in the right-hand side of Eq. (\ref{e16}) are investigated, 
assuming adiabatic conditions for  collapsing quantum Langmuir envelopes. Other scenarios for collapse, 
like the supersonic one \cite{Landman, Papanicolaou}, could also be relevant and shall be investigated in the future. 

\section{Variational solution in two dimensions}

Consider the adiabatic semiclassical system defined by Eq. (\ref{e16}). We refer to localized solution for 
this vector NLS equation as (quantum) ``Langmuir wave packets", or envelopes. As discussed in detail 
in \cite{Gibbons} in the purely classical case, Langmuir wave packets will become singular in a finite time, 
provided the energy is not bounded from below. Of course, explicit analytic Langmuir envelopes are difficult 
to derive. A fruitful approach is to make use of the Lagrangian structure for deriving approximate solutions. 
This approach has been pursued in \cite{Malomed} for the classical and in \cite{Haasvar} for the quantum 
Zakharov system. Both studies considered the internal vibrations of Langmuir envelopes in one-spatial-dimension. 
Presently, we shall apply the time-dependent Rayleigh-Ritz method for the higher-dimensional cases. 
A priori, it is expected that the quantum corrections would inhibit the collapse of localized solutions, 
in view of wave-packet spreading. To check this conjecture, and to have more definite information on 
the influence of the quantum terms, first we consider the following {\it Ansatz},
\begin{equation}
\label{e19}
{\bf\cal{E}} = \left(\frac{N}{\pi}\right)^{1/2}\,\frac{1}{\sigma}\,\exp\left(-\frac{\rho^2}{2\sigma^2}\right)\,
\exp\left(i(\Theta + k\rho^2)\right)\,\,(\cos\phi, \sin\phi, 0) \,,
\end{equation}
which is appropriate for two-spatial-dimensions. Here $\sigma, k, \Theta$ and $\phi$ are real functions 
of time, and $\rho = \sqrt{x^2+y^2}$. The normalization condition (\ref{e12}) is automatically satisfied 
(in 2D the spatial integrations reduce to integrations on the plane). Other localized forms, 
involving {\it e. g.} a {\it sech} type dependence, could have been also proposed. 
Here a Gaussian form was suggested mainly for the sake of simplicity \cite{Fedele}. Notice that the envelope 
electric field (\ref{e19}) is not necessarily electrostatic: it can carry a 
transverse ($\nabla\times{\bf\cal{E}} \neq 0$) component. 

The free functions in Eq. (\ref{e19}) should be determined by extremization of the action functional 
associated with the Lagrangian density (\ref{e17}). A straightforward calculation gives 
\begin{eqnarray}
L_2 \equiv \int\,{\cal L}_{ad,sc}\,dx\,dy &=& - N \,\Bigl[\dot\Theta + \sigma^2 \dot{k} 
+ \frac{2c^2}{v_{Fe}^2}\,k^2 \sigma^2 + \frac{1}{2}\,\left(
\frac{c^2}{v_{Fe}^2} - \frac{N}{2\pi}\right)\,\frac{1}{\sigma^2} \nonumber \\ \label{e20}
&+& 8\Gamma k^2 + 16\Gamma k^4 \sigma^4 + \left(1+\frac{N}{2\pi}\right)\,\frac{\Gamma}{\sigma^4}\,\Bigr] \,,
\end{eqnarray}
where only the main quantum contributions are retained. Now $L_2$ is the Lagrangian for a mechanical system, 
after the spatial form of the envelope electric field was defined in advance via Eq. (\ref{e19}). 
Of special interest is the behavior of the dispersion $\sigma$. For a collapsing solution one could 
expect that $\sigma$ goes to zero in a finite time. The phase $\Theta$ and the chirp function $k$ 
should be regarded as auxiliary fields. Notice that $L_2$ is not dependent on the angle $\phi$, 
which remains arbitrary as far as the variational method is concerned.

Applying the functional derivative of $L_2$ with respect to $\Theta$, we obtain 
\begin{equation}
\label{e21}
\frac{\delta L_2}{\delta\Theta} = 0 \quad \rightarrow \quad \dot{N} = 0 \,,
\end{equation}
so that the variational solution preserves the number of plasmons, as expected. 
The remaining Euler-Lagrange equations are
\begin{eqnarray}
\label{e22}
\frac{\delta L_2}{\delta k} = 0 \quad \rightarrow \quad \sigma\dot\sigma &=& \frac{2 c^2}{v_{Fe}^2}\,\sigma^2 k 
+ 8\Gamma k + 32\Gamma\sigma^4 k^3 \,,\\
\frac{\delta L_2}{\delta\sigma} = 0 \quad \rightarrow \quad \sigma\dot{k} &=& 
- \frac{2c^2}{v_{Fe}^2}\,k^2 \sigma + \frac{1}{2}\,\left(
\frac{c^2}{v_{Fe}^2} - \frac{N}{2\pi}\right)\,\frac{1}{\sigma^3} 
- 32\Gamma k^4 \sigma^3 \nonumber \\ \label{e23} &+& \left(1+\frac{N}{2\pi}\right)\,\frac{2\Gamma}{\sigma^5} \,.
\end{eqnarray}
The exact solution of the nonlinear system (\ref{e22}--\ref{e23}) is difficult to obtain, but at least 
the dynamics was reduced to ordinary differential equations. 

It is instructive to analyze the purely classical ($\Gamma \equiv 0$) case first. This is specially true, 
since to our knowledge the Rayleigh-Ritz method was not applied to the vector NLS equation (\ref{e16}), 
even for classical systems. The reason can be due to the calculational complexity induced by the transverse term. 
When $\Gamma = 0$, Eq. (\ref{e22}) gives $k = v_{Fe}^2 \dot\sigma/2c^2 \sigma$. 
Inserting this in Eq. (\ref{e23}) we have 
\begin{equation}
\label{e24}
\ddot\sigma = - \frac{\partial V_{2c}}{\partial\sigma} \,,
\end{equation}
where the pseudo-potential $V_{2c}$ is 
\begin{equation}
\label{e25}
V_{2c} = \frac{c^2}{2v_{Fe}^2}\,\left(
\frac{c^2}{v_{Fe}^2} - \frac{N}{2\pi}\right)\,\frac{1}{\sigma^2} \,.
\end{equation}
From Eq. (\ref{e25}) it is evident that the repulsive character of the pseudo-potential will be 
converted into an attractive one, whenever the number of plasmons exceeds a threshold,
\begin{equation}
\label{e26}
N > \frac{2\pi c^2}{v_{Fe}^2}  \,, 
\end{equation}
a con\-di\-tion for Lang\-muir wa\-ve pa\-cket co\-llapse in the clas\-si\-cal two-di\-men\-sio\-nal ca\-se.
The interpretation of the result is as follows. When the number of plasmons satisfy Eq. (\ref{e26}), the 
refractive $\sim |{\bf\cal{E}}|^4$ term dominates over the dispersive terms in the Lagrangian density (\ref{e17}), 
producing a singularity in a finite time. Finally, notice the ballistic motion when 
$N = 2\pi c^{2}/v_{Fe}^2$, which can also lead to singularity.

Further insight follows after evaluating the energy integral (\ref{e18}) with the {\it Ansatz} (\ref{e19}), 
which gives, after eliminating $k$,
\begin{equation}
\label{e27}
{\cal H}_{ad,sc,2c} = \frac{N v_{Fe}^2}{c^2}\,\left[\frac{\dot\sigma^2}{2} 
+ V_{2c}\right] \quad (\Gamma \equiv 0) \,.
\end{equation}
Of course, this energy first integral could be obtained directly from Eq. (\ref{e24}). However, the plausibility 
of the variational solution is reinforced, since Eq. (\ref{e27}) shows that it preserves the exact constant 
of motion ${\cal H}_{ad,sc}$. In addition, in the attractive (collapsing) case the energy (\ref{e27}) 
is not bounded from bellow. 

In the quantum ($\Gamma \neq 0$) case, Eq. (\ref{e22}) becomes a cubic equation in $k$, whose exact solution 
is too cumbersome to be of practical use. It is better to proceed by successive approximations, 
taking into account that the quantum and electromagnetic terms are small. In this way, one arrives at 
\begin{equation}
\label{e28}
\ddot\sigma = - \frac{\partial V_{2}}{\partial\sigma} \,,
\end{equation}
where the pseudo-potential $V_{2}$ is 
\begin{equation}
\label{e29}
V_{2} = \frac{c^2}{2v_{Fe}^2}\,\left( 
\frac{c^2}{v_{Fe}^2} - \frac{N}{2\pi}\right)\,\frac{1}{\sigma^2} 
+ \frac{\Gamma c^2}{v_{Fe}^2}\,\left(1+\frac{N}{2\pi}\right)\,\frac{1}{\sigma^4}\,.
\end{equation}
Now, even if the threshold (\ref{e26}) is exceeded, the repulsive $\sim\sigma^{-4}$ quantum term in $V_2$ 
will prevent singularities. This adds quantum diffraction as another physical mechanism, besides dissipation 
and Landau damping, so that collapsing Langmuir wave packets are avoided in vector NLS equation. 
Also, similar to Eq. (\ref{e27}), it can be shown that the approximate dynamics preserves the energy integral, 
even in the quantum case. Indeed, calculating from Eq. (\ref{e18}) and the variational solution 
gives ${\cal H}_{ad,sc}$ as 
\begin{equation}
{\cal H}_{ad,sc,2} = \frac{N v_{Fe}^2}{c^2}\,\left[\frac{\dot\sigma^2}{2} + V_{2}\right] \quad (\Gamma \geq 0) \,.
\end{equation}
From Eq. (\ref{e28}), obviously $\dot{\cal H}_{ad,sc,2} = 0$.

It should be noticed that oscillations of purely quantum nature are obtained when the number of plasmons 
exceeds the threshold (\ref{e26}). Indeed, in this case the pseudo-potential $V_2$ in Eq. (\ref{e29}) 
assumes a potential well form as shown in Figure 1, which clearly admits oscillations around a 
minimum $\sigma = \sigma_{m}$. Here,
\begin{equation}
\label{e30}
\sigma_{m} = 2\,\left[\frac{\Gamma (1 + N/2\pi)}{N/2\pi - c^2/v_{Fe}^2}\right]^{1/2} \,.
\end{equation}
Also, the minimum value of $V_2$ is 
\begin{equation}
V_{2}(\sigma_{m}) = - \frac{c^2}{16\Gamma\,v_{Fe}^2}\,\frac{(N/2\pi - c^2/v_{Fe}^2)^2}{1+N/2\pi} > 
- \frac{1}{16\Gamma}\,\left(\frac{N}{2\pi}-\frac{c^2}{v_{Fe}^2}\right)^2 \,,
\end{equation}
the last inequality follows since Eq. (\ref{e26}) is assumed. Therefore, a deepest potential well 
is obtained when $N$ is increasing. Also, for too large quantum effects the trapping of the localized 
electric field in this potential well would be difficult, since $V_{2}(\sigma_{m}) \rightarrow 0_{-}$ 
as $\Gamma$ increases. This is due to the dispersive nature of the quantum corrections. 

The frequency $\omega$ of the small amplitude oscillations is derived linearizing Eq. (\ref{e28}) around 
the equilibrium point (\ref{e30}). Restoring physical coordinates via Eq. (\ref{e4}) this frequency is calculated as
\begin{eqnarray}
\omega &=& \frac{c}{\sqrt{2}\,v_{Fe}}\,\left(\frac{\kappa_{B}T_{Fe}}{\hbar\,\omega_{pe}}\right)^2\,
\frac{(N/2\pi-c^{2}/v_{Fe}^{2})^{3/2}}{1+N/2\pi}\,\,\omega_{pe} \nonumber \\ \label{e31} 
&<& \frac{v_{Fe}}{\sqrt{2}\,c}\,\left(\frac{\kappa_{B}T_{Fe}}{\hbar\,\omega_{pe}}\right)^2\,
\left(\frac{N}{2\pi}-\frac{c^2}{v_{Fe}^2}\right)^{3/2}\,\omega_{pe} \,.
\end{eqnarray}

\begin{figure*}
\begin{center}
\includegraphics{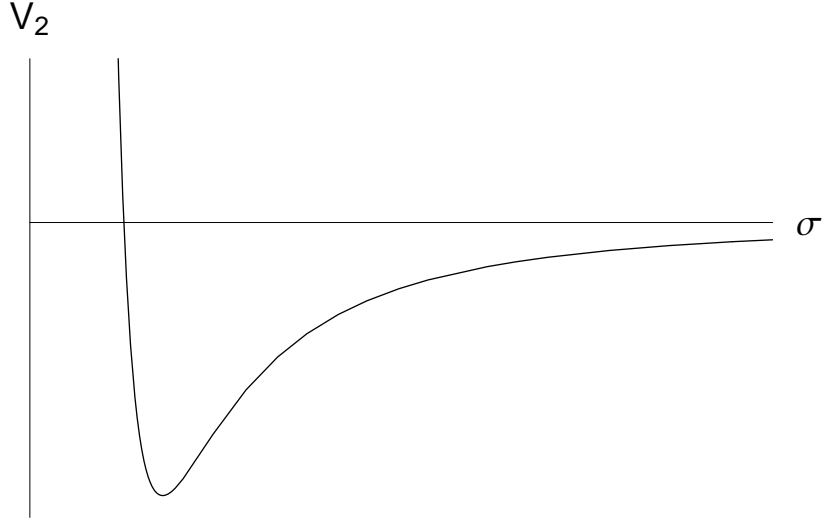}
\caption{The qualitative form of the pseudo-potential in Eq. (\ref{e29}) for $N > 2\pi c^2/v_{Fe}^2$.}
\label{figure1}
\end{center}
\end{figure*}

To conclude, the variational solution suggests that the extra dispersion arising from the quantum terms 
would inhibit the collapse of Langmuir wave packets in two-spatial-dimensions. Moreover, for sufficient 
electric field energy (which is proportional to $N$), instead of collapse there will be oscillations of 
the width of the localized solution, due to the competition between the classical refraction and 
the quantum diffraction. The frequency of linear oscillations is then given by Eq. (\ref{e31}). 
The emergence of a pulsating Langmuir envelope is a qualitatively new phenomena, which could be 
tested quantitatively in experiments. 

\section{Variational solution in three-dimensions}

It is worth to study the dynamics of localized solutions for the vector NLS equation (\ref{e16}) 
in fully three-dimensional space. For this purpose, we consider the Gaussian form 
\begin{equation}
\label{e32}
{\bf\cal{E}} = \left(\frac{N}{(\sqrt{\pi}\,\sigma)^3}\right)^{1/2}\!\!\!\!\!\exp\left[-\frac{r^2}{2\sigma^2}\!
+\!i(\Theta\!+\!k\,r^2)\right](\cos\phi\sin\theta, \sin\phi\sin\theta, \cos\theta) \,,
\end{equation}
where $\sigma, k, \Theta, \theta$ and $\phi$ are real functions of time and $r = \sqrt{x^2+y^2+z^2}$, 
applying the Rayleigh-Ritz method just like in the last Section. The normalization condition (\ref{e12}) 
is automatically satisfied with Eq. (\ref{e32}), which, occasionally, can also support a 
transverse ($\nabla\times{\bf\cal{E}} \neq 0$) part. 

Proceeding as before, the Lagrangian 
\begin{eqnarray}
L_3 \equiv \int\,{\cal L}_{ad,sc}\,d{\bf r} &=& - N \,\Bigl[\dot\Theta + \frac{3}{2}\,\sigma^2 \dot{k} 
+ \frac{4\,c^2}{v_{Fe}^2}\,k^2 \sigma^2 + 
\frac{c^2}{v_{Fe}^2\,\sigma^2} - \frac{N}{4\sqrt{2}\,\pi^{3/2}\,\sigma^3} \nonumber \\ \label{e33}
&+& 10\,\Gamma k^2 + 20\,\Gamma k^4 \sigma^4 + \frac{5\,\Gamma}{4\,\sigma^4}
+\frac{3\,\Gamma N}{4\sqrt{2}\,\pi^{3/2}\,\sigma^5}\,\Bigr] 
\end{eqnarray}
is derived. In comparison to the reduced 2D-Lagrangian in Eq. (\ref{e20}), there are different numerical 
factors as well as qualitative changes due to higher-order nonlinearities. Also, the angular variables $\theta$ 
and $\phi$ don't appear in $L_3$. 

The main remaining task is to analyze the dynamics of the width $\sigma$ as a function of time. This is achieved 
from the Euler-Lagrange equations for the action functional associated to $L_3$. As before, 
$\delta L_{3}/\delta\Theta = 0$ gives $\dot{N}=0$, a consistency test satisfied by the variational solution. 
The other functional derivatives yield 
\begin{eqnarray}
\label{e34}
\frac{\delta L_3}{\delta k} = 0 \rightarrow \sigma\dot\sigma &=& \frac{4k}{3}\,
\left[\frac{2\, c^2}{v_{Fe}^2}\,\sigma^2  + 5\Gamma\,(1+4k^2 \sigma^4)\right]  \,,\\
\frac{\delta L_3}{\delta\sigma} = 0 \rightarrow \sigma\dot{k} &=& \frac{1}{3}\,
\Bigl[- \frac{8\,c^2}{v_{Fe}^2}\,k^2 \sigma + \frac{2\,c^2}{v_{Fe}^2 \sigma^3} 
- \frac{3\,N}{4\,\sqrt{2}\,\pi^{3/2}\,\sigma^4} \nonumber \\ \label{e35} &-& 80\,
\Gamma k^4 \sigma^3 + \frac{5\,\Gamma}{\sigma^5} + \frac{15\,\Gamma N}{4\,\sqrt{2}\,\pi^{3/2}\,\sigma^6}\Bigr] \,.
\end{eqnarray}

In the formal classical limit ($\Gamma \equiv 0$), and using Eq. (\ref{e34}) to eliminate $k$, we obtain 
\begin{equation}
\label{e36}
\ddot\sigma = - \frac{\partial V_{3c}}{\partial\sigma} \,,
\end{equation}
where now the pseudo-potential $V_{3c}$ is 
\begin{equation}
\label{e37}
V_{3c} = \frac{c^2}{v_{Fe}^2}\,\left(
\frac{8\,c^2}{9\,v_{Fe}^2\,\sigma^2} - \frac{2\,N}{9\,\sqrt{2}\,\pi^{3/2}\,\sigma^3}\right) \,.
\end{equation}
The form (\ref{e37}) shows a generic singular behavior, since the attractive $\sim \sigma^{-3}$ term will 
dominate for sufficiently small $\sigma$, irrespective of the value of $N$. Hence, in fully three-dimensional 
space there is more ``room" for a collapsing dynamics. Figure 2 shows the qualitative form of $V_{3c}$, 
attaining a maximum at $\sigma = \sigma_M$, where
\begin{equation}
\label{e38}
\sigma_M = \frac{3\, v_{F}^2\,N}{8\,\sqrt{2}\,\pi^{3/2}\,c^2} \,.
\end{equation}

\begin{figure*}
\begin{center}
\includegraphics{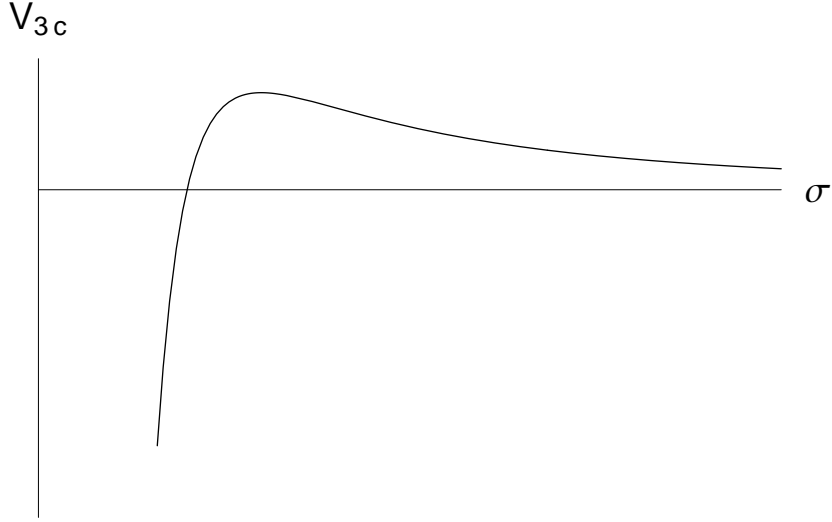}
\caption{The qualitative form of the pseudo-potential $V_{3c}$ in Eq. (\ref{e37}).}
\label{figure2}
\end{center}
\end{figure*}

By Eq. (\ref{e35}) and using successive approximations in the parameter $\Gamma$ to eliminate $k$ via Eq. (\ref{e34}), 
we obtain 
\begin{equation}
\label{ee}
\ddot\sigma = - \frac{\partial V_{3}}{\partial\sigma} \,,
\end{equation}
where 
\begin{equation}
\label{e39}
V_3 = \frac{8\,c^2}{3\,v_{Fe}^2}\,\left[\frac{c^2}{3\,v_{Fe}^2\,\sigma^2} 
- \frac{N}{12\,\sqrt{2}\,\pi^{3/2}\,\sigma^3} + \frac{5\,\Gamma}{12\,\sigma^4} 
+ \frac{\Gamma\,N}{4\,\sqrt{2}\,\pi^{3/2}\,\sigma^5}\right] \,.
\end{equation}
The quantum terms are repulsive and prevent collapse, since they dominate for sufficiently small $\sigma$. 
Moreover, when $\Gamma \neq 0$ an oscillatory behavior is possible, provided a certain condition, to be 
explained in the following, is meet. 

To examine the possibility of oscillations, consider $V_{3}'(\sigma) = 0$, the equation for the critical points 
of $V_3$. Under the rescaling $s = \sigma/\sigma_{M}$, where $\sigma_M$ (defined in Eq. (\ref{e38})) is 
the maximum of the purely classical pseudo-potential, the equation for the critical points read 
\begin{equation}
\label{e40}
V_{3}' = 0 \quad \rightarrow \quad s^3 - s^2 + \frac{4\,g}{27} = 0 \,,
\end{equation}
where 
\begin{equation}
\label{e41}
g = \frac{480\,\pi^3\,\Gamma\,c^4}{N^2\,v_{Fe}^4} 
\end{equation}
is a new dimensionless parameter. In deriving Eq. (\ref{e40}), it was omitted a term negligible except if 
$s \sim c^2/v_{Fe}^2$, which is unlikely. 

The quantity $g$ plays a decisive r\^ole on the shape of $V_3$. Indeed, calculating the discriminant shows 
that the solutions to the cubic in Eq. (\ref{e40}) are as follows: (a) $g < 1 \rightarrow$ three distinct 
real roots (one negative and two positive); (b) $g = 1 \rightarrow$ one negative root, one (positive) double root; 
(c) $g > 1 \rightarrow$ one (negative) real root, two complex conjugate roots. Therefore, $g < 1$ is the condition 
for the existence of a potential well, which can support oscillations. This is shown in Figure 3. 
The analytic formulae for the solutions of the cubic in Eq. (\ref{e40}) are cumbersome and will be omitted. 

\begin{figure*}
\begin{center}
\includegraphics{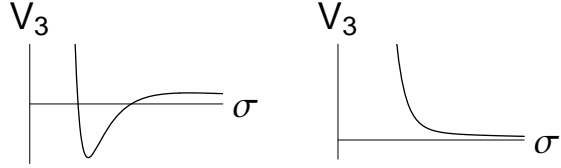}
\caption{The qualitative form of the pseudo-potential $V_{3}$ in Eq. (\ref{e39}) 
for $g < 1$ (on the left) and $g > 1$ (on the right).}
\label{figure3}
\end{center}
\end{figure*}

Restoring physical coordinates, the necessary condition for oscillations is rewritten as
\begin{equation}
\label{e42}
g < 1 \quad \rightarrow \quad \frac{\varepsilon_0}{2}\,\int\,|\tilde{\bf E}|^2\,d{\bf r} > 
\frac{\sqrt{30\pi}}{\gamma}\,\,m_e\,v_{Fe}\,c \,,
\end{equation}
where $\gamma = e^2/4\,\pi\varepsilon_{0}\,\hbar\,c \simeq 1/137$ is the fine structure constant. 
From Eq. (\ref{e42}) it is seen that for sufficient electrostatic energy the width $\sigma$ of the 
localized envelope field can show oscillations, supported by the competition between classical 
refraction and quantum diffraction. Also, due to the Fermi pressure, for large particle densities 
the inequality (\ref{e42}) becomes more difficult to be met, since $v_{Fe} \sim n_{0}^{1/3}$. 
For example, when $n_0 \sim 10^{36}\,m^{-3}$ (white dwarf), the right-hand-side of Eq. (\ref{e42}) 
is $0.6\,$ GeV. For $n_0 \sim 10^{33}\,m^{-3}$ (the next generation intense laser-solid density 
plasma experiments), it is $57.5$ MeV.

Finally, notice that ${\cal H}_{ad,sc}$ from Eq. (\ref{e18}), evaluated with the variational solution (\ref{e32}), 
is proportional to $\dot{\sigma}^2/2 + V_3$, which is a constant of motion for Eq. (\ref{ee}). 
Therefore, the approximate solution preserves one of the basic first integrals of the vector 
NLS equation (\ref{e16}), as it should be. 

\section{Conclusion}

In this paper, the quantum Zakharov system in fully three-dimensional space has been derived. 
An associated Lagrangian structure was found, as well as the pertinent conservation laws. 
From the Lagrangian formalism, many possibilities are opened. Here, the variational description 
was used to analyze the behavior of localized envelope electric fields of Gaussian shape, 
in both two- and three-space dimensions. It was shown that the quantum corrections induce 
qualitative and quantitative changes, inhibiting singularities and allowing for oscillations 
of the width of the Langmuir envelope field. This new dynamics can be tested in experiments. 
In particular, the r\^ole of the parameter $g$ and the inequality in Eq. (\ref{e42}) should be investigated. 
However, the variational method was applied only for the adiabatic and semiclassical case, 
which allows to derive the quantum modified vector NLS equation (\ref{e16}). Other, more general, 
scenarios for the solutions of the fully three-dimensional quantum Zakharov system are also 
worth to study, with numerical and real experiments.  

\vskip .5cm {\bf Acknowledgments} \vskip .5cm

This work was partially supported by the Alexander von Humboldt Foundation. Fernando Haas also thanks 
Professors Mattias Marklund and Gert Brodin for their warm hospitality at the Department of Physics 
of Ume\aa \, University, where part of this work was produced.


\begin{thebibliography}{99}
\bibitem{Zakharov} V. E. Zakharov, Zh. Eksp. Teor. Fiz. {\bf 62}, 1745 (1972) [Sov. Phys. JETP {\bf 35}, 908 (1972)].
\bibitem{Goldman} M. V. Goldman, Rev. Mod. Phys. {\bf 56}, 709 (1984).
\bibitem{Thornhill} S. G. Thornhill and D. ter Haar, Phys. Reports {\bf 43}, 43 (1978).
\bibitem{Garcia} L. G. Garcia, F. Haas, L. P. L. de Oliveira and J. Goedert, Phys. Plasmas {\bf 12}, 012302 (2005). 
\bibitem{Haas} F. Haas, G. Manfredi and M. R. Feix, Phys. Rev. E {\bf 62}, 2763 (2000).
\bibitem{Manfredi} G. Manfredi and F. Haas, Phys. Rev. B {\bf 64}, 075316 (2001).
\bibitem{HaasQMHD} F. Haas, Phys. Plasmas {\bf 12}, 062117 (2005). 
\bibitem{Marklund} M. Marklund, Phys. Plasmas {\bf 12}, 082110 (2005). 
\bibitem{Haasvar} F. Haas, Phys. Plasmas {\bf 14}, 042309 (2007).
\bibitem{Shukla} P. K. Shukla and B. Eliasson, Phys. Rev. Lett. {\bf 96}, 245001 (2006);
Phys. Lett. A {\bf 372}, 2893 (2008).
\bibitem{Tang} X. Y. Tang and P. K. Shukla, Phys. Scripta {\bf 76}, 665 (2007).
\bibitem{Abdou} M. A. Abdou and E. M. Abulwafa, Z. Naturforsch. A {\bf 63}, 646 (2008).
\bibitem{Wakil} S. A. El-Wakil and M. A. Abdou, Nonl. Anal. TMA {\bf 68}, 235 (2008). 
\bibitem{Yang} Q. Yang, C. Q. Dai, X. Y. Wang and J. F. Zhang, J. Phys. Soc. Japan {\bf 74}, 2492 (2005). 
See the comments about this work in Ref. \cite{Tang}.
\bibitem{Misra} A. P. Misra, D. Ghosh and A. R. Chowdhury, Phys. Lett. A {\bf 372}, 1469 (2008).
\bibitem{Zakharov2} V. E. Zakharov, A. F. Mastryukov and V. H. Sinakh, Fiz. Plazmy {\bf 1}, 614 (1975) 
[Sov. J. Plasma Phys. {\bf 1}, 339 (1975)].
\bibitem{Zakharov3} V. E. Zakharov, {\it Handbook of Plasma Physics}, 
eds. M. N. Rosenbluth and R. Z. Sagdeev (Elsevier, New York, 1984), vol. 2, p. 81.
\bibitem{Landman} M. Landman, G. C. Papanicolaou, C. Sulem, P. L. Sulem and X. P. Wang, 
Phys. Rev. A {\bf 46}, 7869 (1992). 
\bibitem{Papanicolaou} G. C. Papanicolaou, C. Sulem, P. L. Sulem and X. P. Wang, 
Phys. Fluids B {\bf 3}, 969 (1991).
\bibitem{Dubois} D. F. Dubois, A. Hanssen, H. A. Rose and D. Russel, J. Geophys. Res. {\bf 98}, 17543 (1993).
\bibitem{Robinson} P. A. Robinson and D. H. Newman, Phys. Fluids B {\bf 2}, 3120 (1990).
\bibitem{Cheung} P. Y. Cheung and A. Y. Wong, Phys. Fluids {\bf 18}, 1538 (1985).
\bibitem{Alinejad} H. Alinejad, P. A. Robinson, I. H. Cairns, O. Skjaeraasen and C. Sobhanian, 
Phys. Plasmas {\bf 14}, 082304 (2007).
\bibitem{Akimoto} K. Akimoto, H. L. Rowland and K. Papadopoulos, Phys. Fluids {\bf 31}, 2185 (1988).
\bibitem{Li} L. H. Li and X. Q. Li, Phys. Fluids B {\bf 5}, 3819 (1993).
\bibitem{Pelletier} G. Pelletier, H. Sol and E. Asseo, Phys. Rev. A {\bf 38}, 2552 (1988). 
\bibitem{Stenflo} L. Stenflo, Phys. Rev. Lett. {\bf 48}, 1441 (1982). 
\bibitem{Marklund2} M. Marklund, G. Brodin and L. Stenflo, Phys. Rev. Lett. {\bf 91}, 163601 (2003).
\bibitem{Gibbons} J. Gibbons, S. G. Thornhill, M. J. Wardrop and D. ter Haar, J. Plasma Phys. {\bf 17}, 153 (1977).
\bibitem{Malomed} B. Malomed, D. Anderson, M. Lisak, M. L. Quiroga-Teixeiro and L. Stenflo, 
Phys. Rev. E {\bf 55}, 962 (1997).
\bibitem{Fedele} R. Fedele, U. de Angelis and T. Katsouleas, Phys. Rev. A {\bf 33}, 4412 (1986).
\end{thebibliography}
\end{document}